# EFFECTS OF DARK MATTER IN STAR FORMATION

**Kenath Arun**

Department of Physics and Electronics, Christ (Deemed to be University), Bengaluru-560029, Karnataka, India, &

Department of Physics, Christ Junior College, Bengaluru-560029, Karnataka, India

e-mail: kenath.arun@cjc.christcollege.edu

**S B Gudennavar**

Department of Physics and Electronics, Christ (Deemed to be University), Bengaluru-560029, Karnataka, India

e-mail: shivappa.b.gudennavar@christuniversity.in

**A Prasad**

Center for Space Plasma & Aeronomic Research, The University of Alabama in Huntsville, Huntsville, Alabama 35899

email: ap0162@uah.edu

**C Sivaram**

Indian Institute of Astrophysics, Bengaluru-560034, Karnataka, India

e-mail: sivaram@iiap.res.in

**Abstract:** The standard model for the formation of structure assumes that there existed small fluctuations in the early universe that grew due to gravitational instability. The origins of these fluctuations are as yet unclear. In this work we propose the role of dark matter in providing the seed for star formation in the early universe. Very recent observations also support the role of dark matter in the formation of these first stars. With this we set observable constraints on luminosities, temperatures, and lifetimes of these early stars with an admixture of dark matter.

# 1. Introduction

In a recent paper (Arun et al., 2018) that discussed the effects of admixture of dark matter (DM) particles in white dwarfs, it was shown that this admixture can lead to lowering of their inherent luminosity and as a consequence explain away the need for dark energy (DE). At earlier epochs, the density of DM particles would have been higher, hence we also considered the possibility of WD accreting these DM particles. Here we look at the possibility of stars in the early universe having an admixture of DM particles along with the regular baryonic matter, since at early epoch the density of DM particles would be higher.

The standard cosmological model describe the evolution of the universe following the big bang. The observations from the cosmic microwave background shows that the early universe was smooth, with evidence of small-scale density fluctuations. The cosmological models predict that these clumps would gradually evolve into gravitationally bound structures. Smaller systems would form first and then merge into larger ones (Barkana and Loeb, 2001). These regions at early epoch would have a higher density of DM particles, and hence contribute to star formation in the early universe.

Very recent observations (Bowman et al., 2018) have determined that the earliest stars formed just 180 million after the big bang, predating the earlier oldest stars observed to form around 400 million years after the big bang. This study observed the altered excitation state of 21-centimetre hyperfine line as the light from the first stars penetrate the primordial hydrogen gas. But the observation that the signal is twice that was expected indicate that the hydrogen gas was significantly cooler. This could possibly be an indication of DM (Barkana, 2018), making this the first observation of DM other than through its gravitational effects.

In this context it is of interest that in a recent paper (Sivaram, Arun and Kiren, 2018) we have pointed out in connection with the formation of supermassive black holes in the early universe, that a cloud of DM dominated matter in a star forming region of radius of 200pc at high redshift ($z$ ~50) would have collapsed on a time scale of $10^8$ years. In fact we have stated that this collapsing gas could form clusters of massive stars (blue giants etc.). This is of the same timescale as the recent observation.

So our main idea is that the ambient density of DM particles (being significantly higher at earlier epoch) would have affected the physical properties of the earliest objects to form, such as primordial stars. We assume that the DM particles interact only gravitationally with ordinary matter (and among themselves) and are not coupled to radiation. So the additional gravitational energy would heat up only the baryonic matter. DM also would not contribute to optical opacity, and this affects the luminosity and lifetimes of these primeval stellar objects.

## 2. Dark matter objects

We consider gravitationally bound objects made of DM matter particles, with the DM particle mass varying form 10 – 100 GeV (Narain, Schaffner-Bielich and Mishustin, 2006; Sivaram and Arun, 2011). These particles are assumed to be CDM particles and are also assumed to be fermions. Only in this case would the degeneracy pressure be important which follows the arguments in an earlier paper (Sivaram and Arun, 2011). These objects will have low non-thermal energies and hence the degeneracy pressure will be dominant. The Chandrasekhar mass for such objects (which will be the upper limit on their mass) is given by:

$$M_{ch(DM)} = \left(\frac{\hbar c}{G}\right)^{3/2} \frac{1}{m_D^2} \qquad (1)$$

where, $m_D$ is the mass of the dark matter particle. Currently DM particle of mass, $m_D \sim 60 GeV$ is favoured from results like DAMA experiment (Gelmini, 2006). There is also interest in the detection of excess gamma-rays from the galactic centre, which is attributed to the decay of 60 GeV DM particles (Huang, Zhang and Zhou, 2016). For a review on various models of DM see (Arun, Gudennavar and Sivaram, 2017). For such 60 GeV DM particles this Chandrasekhar mass limit works out to be:

$$M_{ch(DM)} \approx 10^{30} g \approx 10^{-3} M_\odot \qquad (2)$$

The corresponding size of these objects is given by (for the usual degenerate gas configuration) (Sivaram and Arun, 2011):

$$M_{ch(DM)}^{1/3} R = \frac{92\hbar^2}{G m_D^{8/3}} \qquad (3)$$

For the $10^{-3} M_\odot$ object the size works out to be:

$$R \approx 10^5 cm \tag{4}$$

The above discussions are for pure DM objects. Next we explore the formation and evolution of stellar structures with varying proportion of DM admixed with ordinary baryonic matter.

## 3. Possible scenario for star formation with DM

The discussion in section 2 applies to the final state of these DM objects. We now look at the scenario where the primordial mixture of dark matter and baryonic matter is contracting to form the early stars. Early galaxies have been detected at a redshifts of about 10, which indicates that the collapse to form the early stellar structures would have happened even earlier, say at a $z \sim 20$ (van den Bosch, 2001; Firmani and Avila-Reese, 2003). To estimate the ambient density of the DM particles at that epoch, we have to first calculate the critical density and assume that DM constitute ~0.25 of this critical density, given by,

$$\rho_C = 3H^2/8\pi G \tag{5}$$

Using $H$ ~68km/s/Mpc, as implied by the latest Planck observations (Ade et al., 2016), we get the present critical density as, $\rho_0 \approx 10$ keV/cm$^3$, and the present energy density of DM as, $\rho_{DM(0)} = 0.25 \times 10$ keV/cm$^3$. The density of the DM particles at a particular redshift $z$ is given by:

$$\rho_{DM(z)} = 0.25 \rho_0 (1+z)^3 \tag{6}$$

For DM particles of mass $m_D = 60$ GeV, the number density of dark matter particles at $z \sim 20$ is given by:

$$n = 0.25 \left(\frac{10 keV}{60 GeV}\right)(1+20)^3 \approx 10^{-3}/cc \tag{7}$$

Equation (7) gives the ambient density of $10^{-3}$ particles/cc. Now different models (for e.g. spherical top-hat model) of structure formation require the density of the cloud after collapse to be at least 100-200 times the ambient (interstellar) density. Here we assume a minimum density to be ~200 times the ambient density. This gives, $n_{cloud} = 0.2/cc$. So that the mass density of such a collapsing cloud of DM particles (of mass $m_D = 60$ GeV, and at $z \sim 20$) is then given by:

$$\rho_{DM} \approx \left(10^{-22} g\right)\left(2 \times 10^{-1} / cc\right) = 2 \times 10^{-23} g/cc \qquad (8)$$

In addition to DM, we now have baryonic matter also present in the collapsing cloud. Indeed the number of DM particles present is assumed to be a fraction '$f$' of the number of baryonic matter particles. Hence the number density of DM particles can be written in terms of the baryonic number density as $n_{DM} = fn_B$, and the corresponding mass densities are $\rho_{DM} = n_{DM}m_{DM}$, $\rho_B = n_B m_B$. The number density of baryonic matter is 10 to 100 times that of DM particles, but the DM particle mass is ~60 times that of the baryonic particle's mass. The total density is:

$$\rho = \rho_{DM} + \rho_B = n_{DM}m_{DM} + n_B m_B \qquad (9)$$

Now the time taken for the initial cloud (of DM and baryonic matter) to collapse is:

$$t_{collapse} \sim \frac{1}{\sqrt{G\rho}} \approx 10^{15} s \qquad (10)$$

As the cloud collapses, the DM particles are not coupled with radiation whereas the baryonic matter will be heated up, i.e. the presence of DM particles increases the gravitational energy (given by $GM_T^2/R$, where $M_T$ is the total mass), but only baryonic matter gets heated (given by $M_B R_g T$, where $M_B$ is the baryonic mass). Hence we have:

$$M_B R_g T - \frac{GM_T^2}{R} \leq 0 \qquad (11)$$

where $R_g$ is the universal gas constant, and the total mass, $M_T = M_D + M_B = (1+f)M_B$, as before the number of DM particles present is assumed to be a fraction '$f$' of the number of baryonic matter particles, i.e. $M_D = fM_B$; $f \leq 1$. Hence the temperature of this baryonic matter is given by:

$$T = (1+f)^2 \frac{GM_B}{R_g R} \qquad (12)$$

where the baryonic mass is about ten times that of the DM core (as given by equation (2)), i.e. $M_B = 10^{31} g$, and the corresponding degenerate size of $R \approx 10^9 cm$. For an object of equal amount of DM and baryonic matter, i.e. $f = 1$, the temperature works out to be, $T \approx 10^7 K$.

The energy density due to radiation will be comparable to the collapsing gas's energy density. This is true in the stars since if the radiation density exceeds the gas density, instability can set in as is well known, so in the limiting case, when the two are comparable, we have (in general it could be some fraction of the order of 1):

$$aT^4 = nk_B T \tag{13}$$

For $T \approx 10^7 K$, the number density will be:

$$n \approx 10^{22} /cc \tag{14}$$

This is the number density at the stellar cores. At this temperature (equation (12)) and densities (equation (14)), thermonuclear reactions will start. This scenario of baryonic matter collapsing along with the DM particles could thus lead to the formation of stars. There will be gamma ray emissions as the thermonuclear reaction proceeds in the star. But these will be MeV gamma rays as opposed to the GeV gamma rays coming from the DM objects. The possible masses of these objects can be estimate from the energy released during the collapse, which is given by:

$$E_{grav} = \frac{GM_B M_D}{R} \approx 10^{42} \, ergs \tag{15}$$

This gravitational potential energy released will heat up the baryonic matter accreting on to the objects (DM would not be thermally coupled at all):

$$M_B R_g T = 10^{42} \, ergs \tag{16}$$

where $M_B$ is the baryonic mass, and from equation (12), temperature $T \approx 10^7 K$. The mass (baryonic) of these objects is therefore, $M_B = 10^{27} g$, which is sub-stellar mass, and the total mass is, $M_T = M_D + M_B$. Giving a volume of, $V = \frac{M}{nm_D} \approx 10^{27} cc$, and radius of $\sim 10^9 cm$.

This could give rise to a new class of stellar objects, with the DM core of mass $10^{30} g$ and size $\sim 10^5 cm$ surrounded by a layer of collapsed baryonic matter. For bound structure star formation we have the condition given by equation (11) as:

$$M_B R_g T - \frac{GM_T^2}{R} \leq 0 \tag{17}$$

Where again we have: $M_T = (1+f)M_B$, then equation (17) gives:

$$M_B R_g T = (1+f)^2 \frac{GM_B^2}{R} \tag{18}$$

The size of the object is then given by:

$$R = \left( \frac{R_g T}{G(1+f)^2 \frac{4}{3}\pi\mu\rho_B} \right)^{1/2} \tag{19}$$

Where, $M_B = \mu n_B m_p = \mu\rho_B$; $\mu$ is the molecular mass ($\mu = 1$ for a cloud of hydrogen), $n_B$ baryon number density, $m_p$ the proton mass and $\rho_B$ baryon mass density. And the corresponding mass is:

$$M_T = \frac{4}{3}\pi R^3 \mu\rho_B (1+f) \tag{20}$$

For a 10% DM admixture in a pure hydrogen cloud, the size and mass is of the order of $R = 1.3 \times 10^{19}$ cm and $M_T \approx 10^{36}$ g. As the fraction of DM particles increases the size and mass decreases, as can be seen from figures 1 and 2, and as a result the star formation in such a scenario is easier.

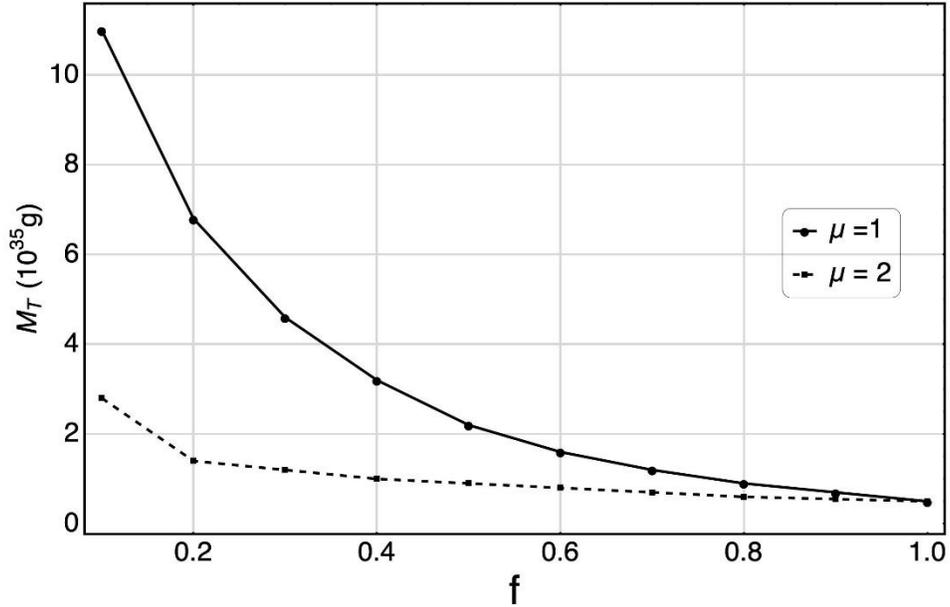

Figure 1: Change in mass with increasing DM fraction

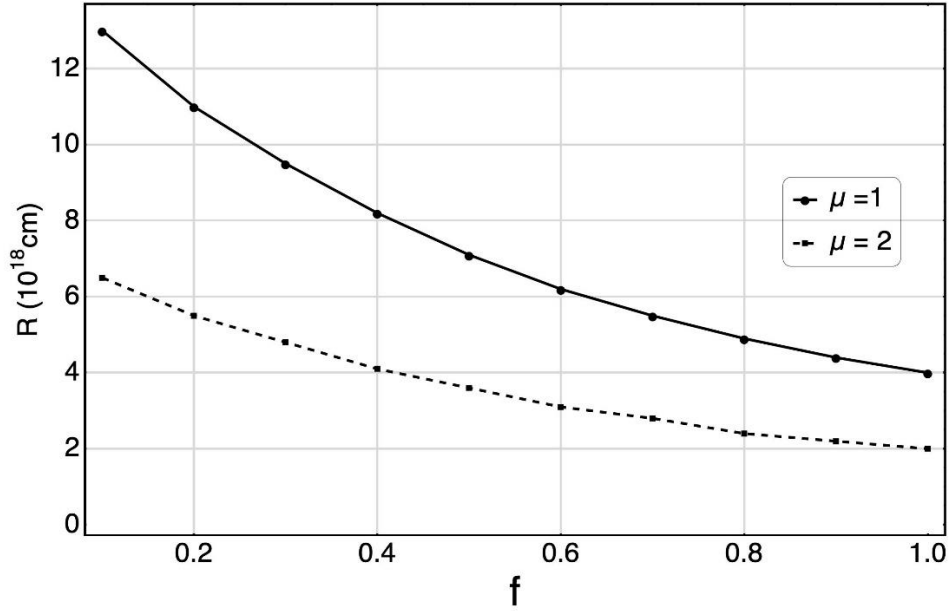

Figure 2: Change in radius with increasing DM fraction

If the star forming cloud has an angular velocity of $\omega$, the rotational energy will contribute against gravity. Therefore we have:

$$M_B R_g T - \frac{GM_T^2}{R} + \frac{2}{5} M_T R^2 \omega^2 \leq 0 \qquad (21)$$

The angular momentum, $M_T R^2 \omega = J$, of the cloud is a constant. The angular momentum of the star forming cloud will be of the order of that of the solar system, i.e. $\sim 100\, J_\odot = 10^{52}\, g\, cm^2\, s^{-1}$. For a Keplerian rotation, the size of the cloud is given by:

$$R = \left( \frac{R_g T}{\frac{4}{3}\pi(1+f)^2 G\rho_B - \frac{2}{5}(1+f)\omega^2} \right)^{1/2} \qquad (22)$$

For a DM fraction of, $f \sim 1$ (i.e. baryonic matter and DM in equal part), $R \sim 10^{19} cm$ and mass, $M_T \sim 5 \times 10^{35} g$. The size and the corresponding mass is of similar order as that without rotation of the cloud.

The star forming cloud could also have a magnetic field ($B$) that could affect the star formation by preventing complete collapse of the cloud to produce stars. Including the effects of magnetic field, for the cloud to collapse, we should have the energy equation as:

$$M_B R_g T - \frac{GM_T^2}{R} + \frac{2}{5}M_T R^2 \omega^2 + \frac{B^2}{8\pi}\left(\frac{4}{3}\pi R^3\right) \leq 0 \quad (23)$$

The effects of the magnetic field will be comparable to that of thermal energy for a magnetic field given by:

$$\frac{B^2}{8\pi} = \rho_B R_g T \Rightarrow B \approx 10^{-6} G \quad (24)$$

These effects of magnetic field will be studies by the infrared camera called the High-resolution Airborne Wideband Camera-Plus (HAWC+), installed on the Stratospheric Observatory for Infrared Astronomy (SOFIA). For the magnetic effect to be comparable to gravitation, we have:

$$\frac{GM^2}{R} = \frac{B^2}{8\pi}\left(\frac{4}{3}\pi R^3\right) \quad (25)$$

where $M = \rho\left(\frac{4}{3}\pi R^3\right)$. This gives a size of $R \approx 10^{19} cm$ which is about 10 light-years, and a corresponding mass of $M \approx 200\, M_\odot$. From simulations of a magnetised, collapsing region in which radiative feedback occurs, Price and Bate (2009) concluded that a strong magnetic field and radiative feedback leads to a star-formation rate of less than about 10% per free-fall time. But the admixture of DM could help in enhancing the star formation rate.

## 4. Effect of DM on stellar luminosities, temperature, and lifetimes

The temperature is related to the luminosity by the radiative transport equation as:

$$T^3 \frac{dT}{dr} = \frac{3\kappa\rho}{4ac}\frac{L(r)}{4\pi r^2} \quad (26)$$

where, $L(r)$ is the luminosity, $T$ is the temperature, $\kappa\rho$ is the cross section. From equation (12), the temperature is given by, $T = (1+f)^2 \frac{GM_B}{R_g R}$. The density is given by $\rho = \frac{M_T}{R^3} = (1+f)\frac{M_B}{R^3}$. Using these in equation (26) we have:

$$(1+f)^8 \frac{G^4 M_B^4}{R_g^4 R^5} = (1+f)\frac{M_B}{R^3}\frac{3\kappa}{4ac}\frac{L}{4\pi R^2} \quad (27)$$

As typical examples, we consider the cases of Thompson and Kramers' opacity. In the case of high mass (hot) stars the opacity is determined by Thompson scattering $\sigma_T$. For a DM admixture fraction of $f$, the luminosity increases by a factor of $(1+f)^7$.

For a lower mass, late-type main sequence star, the opacity is governed by Kramers' law, where the opacity is given by: $\kappa = \kappa_0 \rho T^{-7/2}$. Hence from equations (26) and (27) it follows that for a DM admixture fraction of $f$, the luminosity will increase by a factor of $(1+f)^{15}$. The variation of luminosities to luminosities without DM ($L_0$) in these cases are given in figure 3.

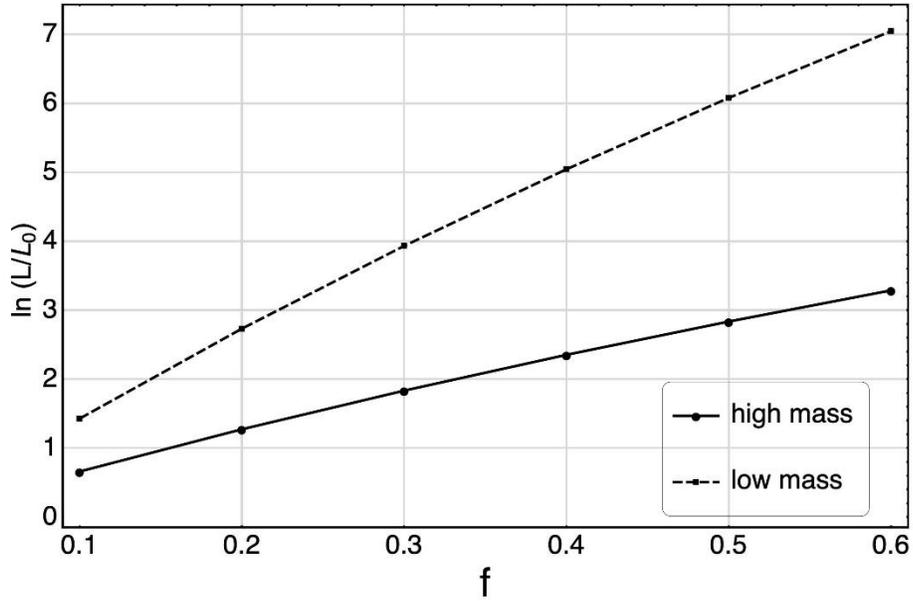

Figure 3: Change in luminosity with increasing DM fraction

As expected, more the fraction of DM particles, the main sequence stars are more luminous, since now the gravitational energy is more, heating up the baryonic matter to higher temperatures. In general the core temperature will increase. From equation (18), we have:

$$T = (1+f)^2 \frac{GM_B}{R_g R} \qquad (28)$$

A DM faction of $f \sim 0.1$, will lead to a core temperature increase of 1.2, and since nuclear reactions are highly sensitive to temperature, massive stars could be significantly more luminous. In the absence of any DM constituents, the core temperature is given as $T_0 = \dfrac{GM_B}{R_g R}$. The ratio of the core temperature with increasing DM fraction to $T_0$ is plotted in figure 4.

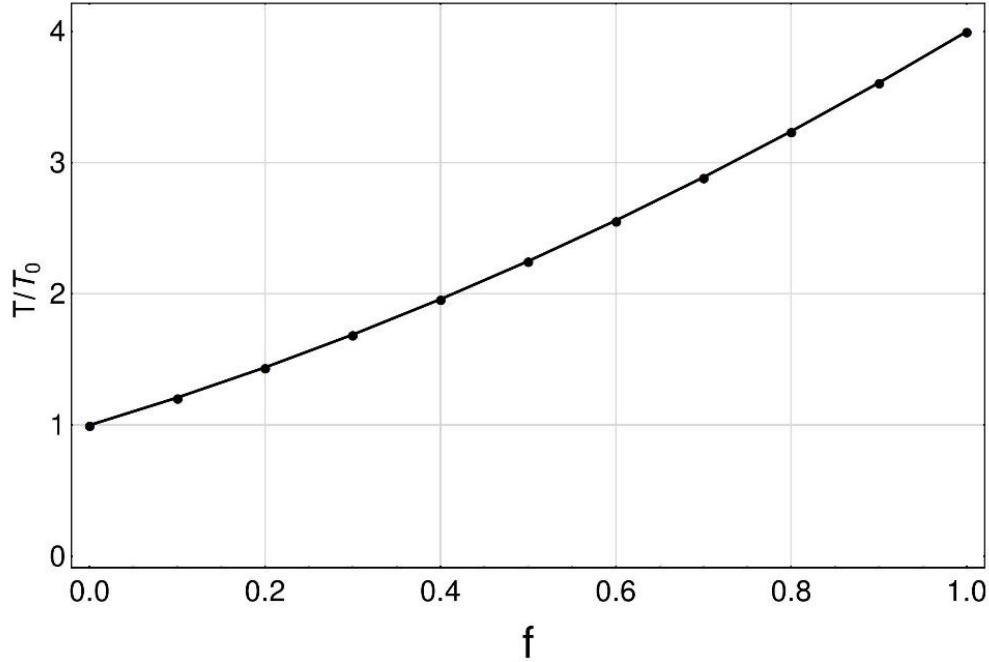

Figure 4: Change in core temperature with increasing DM fraction

The Eddington luminosity will change with the change with the fraction of DM particles. Gravity is determined by both baryonic and DM particles, but radiation pressure opposing gravity is due to only baryonic matter. The radiation pressure is given by $\dfrac{L}{4\pi R^2 c}\sigma_T$, and gravitational pressure is given by: $\dfrac{GM_T m_p}{R^2}$, hence the maximal (Eddington) luminosity is given as:

$$L_{max} = \dfrac{4\pi GM_B(1+f)m_p c}{\sigma_T} \qquad (29)$$

where, $M_T = (f+1)M_B$, so for a DM fraction, $f \sim 0.1$, $L_{max} = 1.1 L_{Edd}$, and $f \sim 1$ (i.e. equal amounts of DM and baryonic matter), $L_{max} = 2L_{Edd}$.

This could also affect the luminosities of quasars. If DM clouds are collapsing along with gas (baryonic) onto supermassive black holes in quasars, their maximal luminosity could be higher. This would imply corrections to the masses of SMBH estimated from their luminosity (assumed to be maximal).

Since the luminosity of stars is increased by the presence of DM particles, their lifetimes will be correspondingly reduced with an increase in the admixture of DM particles. The mass available for nuclear reactions is the baryonic mass whereas the luminosity is affected by DM admixture. The Salpeter lifetime for objects emitting at maximal Eddington luminosity is now given by:

$$\tau_{life} = \frac{M_B c^2}{L_{max}} = \frac{c\sigma_T}{4\pi G(1+f)m_p} \qquad (30)$$

A 10% admixture of DM particles can result in the lifetime decreasing to about 90% of the lifetime of the same stars without any DM particles. In the case of O-type main sequence stars, the lifetime is:

$$\tau_{life} = \frac{\eta M_B c^2}{L_{max}} \qquad (31)$$

where $\eta = 0.007$ is the thermonuclear efficiency. The lifetimes is lowered by a factor of $(1+f)$, as given by equation (30). For a DM fraction of about 20%, their lifetimes could be reduced to $\sim 3\times 10^6$ years, from their typical lifetimes of about 5 – 6 million years as given by equation (31).

As discussed earlier, in the case of high mass stars, the luminosity increases by a factor of $(1+f)^7$. And for a lower mass, late-type main sequence star, the correspondingly the luminosity increase by a factor of $(1+f)^{15}$. As a result the lifetimes of high mass and low mass stars will be lowered by a factor of $(1+f)^7$ and $(1+f)^{15}$, respectively. A 10% DM admixture will result in the lifetime of high mass stars decreasing by a factor of ~2 and of low mass stars by a factor of ~4. The variation of lifetimes with increasing DM fraction to the lifetimes without any DM admixture ($\tau_0$) is plotted in figure 5.

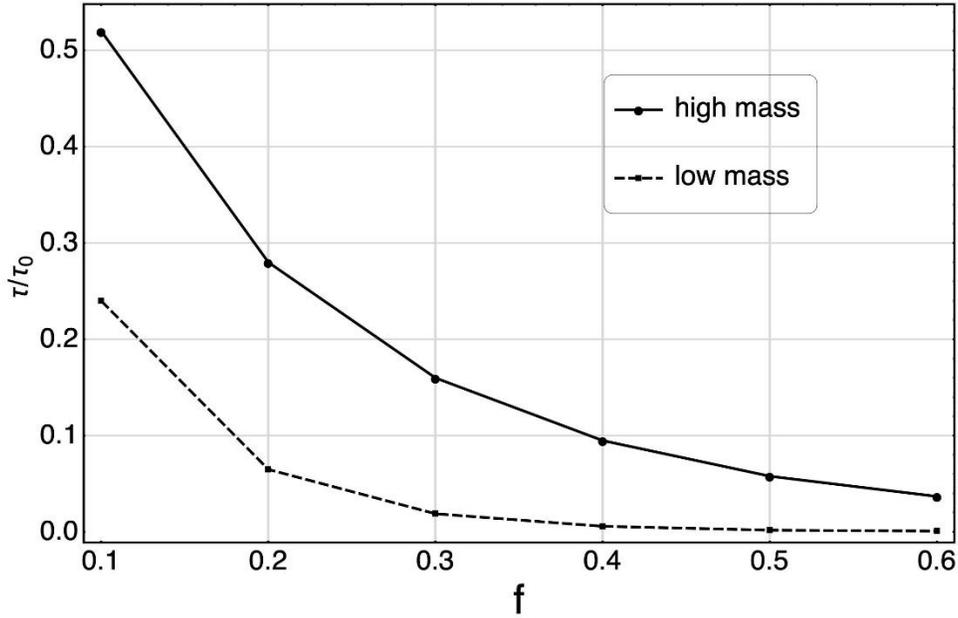

Figure 5: Change in lifetime with increasing DM fraction

## 5. Summary and Conclusions

The formation of the earliest stellar structures is still poorly understood. Even the most recent observations of the absorption profile centred at 78MHz in the sky-averaged spectrum pushes the cosmic dawn to just 180 million years after the big bang. Here we have proposed the role of dark matter in providing the seed for star formation in the early universe. As the density of DM particles at earlier epochs was higher, this scenario is a possibility. As stated in the introduction, this is consistent with the recent observation. As can be seen from figures 1 and 2, the admixture of DM with baryonic matter reduces the Jean's mass and size and hence providing an easier scenario for star formation. The admixture of DM would affect the properties of these early stars such as their luminosities, core temperatures, and lifetimes, as can be seen from figures 3, 4, and 5. These plots provide the variation in these properties with increasing fraction of DM particles admixed with baryonic matter in the early stars.

# References

- Ade, P.A.R. et al., 2016. Planck 2015 results XIII. Cosmological parameters. Astron. Astrophys. 594, A13.
- Arun, K., Gudennavar, S.B., Sivaram, C., 2017, Dark matter, dark energy, and alternate models: A review. Adv. Space Res., 60, 166.
- Arun, K. et al., 2018. Alternate Models to Dark Energy. Adv. Space Res., 61, 567.
- Barkana, R., 2018. Possible interaction between baryons and dark-matter particles revealed by the first stars. Nature, 555, 71.
- Barkana, R., Loeb, A., 2001. In the beginning: the first sources of light and the reionization of the universe. Phys. Rep., 349, 125.
- Bowman, J.D. et al., 2018. An absorption profile centred at 78 megahertz in the sky-averaged spectrum. Nature, 555, 67.
- Firmani, C., Avila-Reese, V., 2003. Physical processes behind the morphological Hubble sequence. Rev. Mex. Astron. Astrofis., 17, 107.
- Gelmini, G.B., 2006. DAMA detection claim is still compatible with all other DM searches. J. Phys. Conf. Ser., 39, 166.
- Huang, X-J., Zhang, W-H., Zhou, Y-F., 2016. 750 GeV diphoton excess and a dark matter messenger at the Galactic Center. Phys. Rev. D, 93, 115006.
- Narain, G., Schaffner-Bielich, J., Mishustin, I.N., 2006. Compact stars made of fermionic dark matter. Phys. Rev. D, 74, 063003.
- Price, D.J., Bate, M.R., 2009. Inefficient star formation: the combined effects of magnetic fields and radiative feedback. Mon. Not. R. Astron. Soc., 398, 33.
- Sivaram, C., Arun, K., 2011. New Class of Dark Matter Objects and their Detection. Open Astron. J., 4, 57.
- Sivaram, C., Arun, K., Kiren, O.V., 2018. Forming supermassive black holes like J1342+0928 (invoking dark matter) in early universe. Astrophys. Space Sci., 363, 40.
- van den Bosch, F.C., 2001. The origin of the density distribution of disc galaxies: a new problem for the standard model of disc formation. Mon. Not. R. Astron. Soc., 327, 1334.